\documentclass[12pt]{nature}

\usepackage{scicite}
\usepackage{times}
\usepackage{amsmath}
\usepackage{graphicx}
\usepackage{afterpage}
\usepackage{txfonts}
\usepackage{color}
\usepackage[version=4]{mhchem}

\newenvironment{sciabstract}{%
\begin{quote} \bf}
{\end{quote}}

\title{Miniaturized spectrometer enabled by end-to-end deep learning on large-scale radiative cavity array}

\author
{Xinyi Zhou$^{1}$, Cheng Zhang$^{1,2}$, Xiaoyu Zhang$^{1}$, Yi Zuo$^{1}$, Zixuan Zhang$^{1}$, Feifan Wang$^{1}$, Zihao Chen$^{1}$, Hongbin Li$^{1,2}$, Chao Peng$^{1,2\ast}$\\
\\
\normalsize{$^{1}$State Key Laboratory of Advanced Optical Communication Systems and Networks, School of Electronics \& Frontiers Science Center for Nano-optoelectronics, Peking University, Beijing 100871, China}\\
\normalsize{$^{2}$Peng Cheng Laboratory, Shenzhen 518055, China}\\
\normalsize{$^\ast$To whom correspondence should be addressed; E-mail: Chao Peng (pengchao@pku.edu.cn) }\\\\
}

\date{}

\begin{document} 


\baselineskip 24pt


\maketitle

\begin{sciabstract}
Miniaturized (mini-) spectrometers are highly desirable tools for chemical, biological, and medical diagnostics because of their potential for portable and in situ spectral detection. In this work, we propose and demonstrate a mini-spectrometer that combines a large-scale radiative cavity array with end-to-end deep learning networks. Specifically, we utilize high-$Q$ bound states in continuum cavities with distinct radiation characteristics as the fundamental units to achieve parallel spectral detection. We realize a $36\times30$ cavity array that spans a wide spectral range from $1525$ to $1605$ nm with quality factors above $10^4$. We further train a deep network with $8000$ outputs to directly map arbitrary spectra to array responses excited by the out-of-plane incident. Experimental results demonstrate that the proposed mini-spectrometer can resolve unknown spectra with a resolution of $0.048$ nm in a bandwidth of $80$ nm and fidelity exceeding $95\%$, thus offering a promising method for compact, high resolution, and broadband spectroscopy.
\end{sciabstract}

\section*{Introduction}

Spectroscopy is a vital method in characterizing the spectral features of light, and thus can find applications spanning chemical\cite{2004bioche}, agriculture\cite{2019agriculture}, medicine\cite{2023medicine,2023medicine-roman}, and communications\cite{2020laboratory-comminication}. Due to their unique capability of portable and in situ diagnosis\cite{2021yang-miniaturization,li2022advances,tan2023optical,kim2023drug}, miniaturized (mini-) spectrometers have become a burgeoning field with significant advances by applying dispersive\cite{2007disp,2015disp,2019disp,2022disp,2022disp1,2023disp} or narrow-band optical components\cite{2000NB,2010NB,2016FP,2023phc} and materials\cite{yuan2021wavelength,zheng2022single,wang2023room,deng2022electrically,yoon2022miniaturized}, Fourier transform interferometers\cite{2017fouri,2018fourier,2018fourier-1,2019fourier,2019fourier-1,2020fouri,2024fourier}, and computational spectral reconstruction\cite{2019phc,yang2023deep,2019nanowire,cheng2021generalized,2021quantum,gao2022computational,bian2024broadband,wang2007concept}. In principle, photons can be spectrally distinguished by sufficiently different light paths or lifetimes. Since the limited volume in mini-spectrometers restricts their ability to extend the light path, and the narrow-band components struggle to cover the entire desired spectrum, achieving the optimal size, detection resolution, and range in mini-spectrometers involves complex trade-offs. However, highly spectrally selective components, such as high-$Q$ optical resonators, are always essential and advantageous to achieve high resolution\cite{2023phc,2023L3array,xu2023integrated,2019microring,yao2023broadband,zhang2022ultrahigh}. By arranging high $Q$ resonators in an array manner, the detection range can be expanded while keeping the devices' size compact using state-of-the-art nanofabrication technology. Consequently, various optical resonators, such as nanowires\cite{2021nanowire,2019nanowire}, quantum dots\cite{2015quantum,2020quantum,2021quantum}, microring resonators\cite{2019microring,2008cavitymicroring,xu2023integrated,yao2023broadband}, photonic crystals\cite{2019phc,2023phc}, and plasmonic metasurfaces\cite{2021plasm,2021plsma,2023plasm,tang2024metasurface,zhang2024miniature}, are employed for parallel spectrum detection. In order to achieve high resolution and a broad detection range, the resonator array must be large-scale, which presents significant challenges in connecting, scanning, and calibrating these components. For this purpose, ultra-high-$Q$ radiative optical cavities capable of out-of-plane excitation could be a promising candidate, since they can avoid the need for planar connections and temporal scanning as a micro-ring resonator array does. Recently, a new type of optical cavity has been reported utilizing miniaturized bound states in the continuum (mini-BICs) with quality factors ($Q$s) above 1 million in a size of $20~\mu$m\cite{2022phcchen}. Such mini-BIC cavities possess a set of discrete optical modes with distinct radiation features in the out-of-plane direction, thus showing a suitable primitive array unit for a mini-spectrometer. The calibration difficulties can be relaxed by applying end-to-end machine learning (ML) techniques\cite{2019end,2020end,2021end,chen2019end,zhang2020solver}, that is, the correspondence between the output of the matrix and the features of the spectrum can be trained by neural networks, without one-by-one calibrating the response of individual cavities. By integrating the aforementioned techniques, we propose that a compact, high-resolution, and broadband mini-spectrometer is feasible from the array arrangement of radiative cavities combined with end-to-end deep learning techniques.

In this work, we propose and demonstrate an on-chip miniaturized spectrometer that leverages the array-arranged mini-BIC cavities on photonic crystal (PhC) slabs to detect comprehensive and sophisticated spectral features of lights with high resolution and broadband. Briefly, the cavities show high $Q$s and rich radiative patterns under out-of-plane excitation of incident light. We carefully optimize the arrangement of the array of 1080 ($30\times 36$) cavities on a footprint of $505~\mu$m $\times$ $606~\mu$m to diversify their radiative characteristics and promote their coverage in the spectrum. By employing an end-to-end deep learning techniques, we train a neural network that effectively maps the array responses to spectral features. We experimentally show that the mini-spectrometer is capable of reconstructing arbitrary blind spectra in a range of $1525- 1605$ nm, with a minimal distinguishable resolution of $0.048$ nm and detection sensitivity of $12.5~\mu$W/nm, showing a promising method for realizing compact, high-performance on-chip spectrometers.

\section*{Principle and design}

\begin{figure}[ht]
\centering\includegraphics[width=18cm]{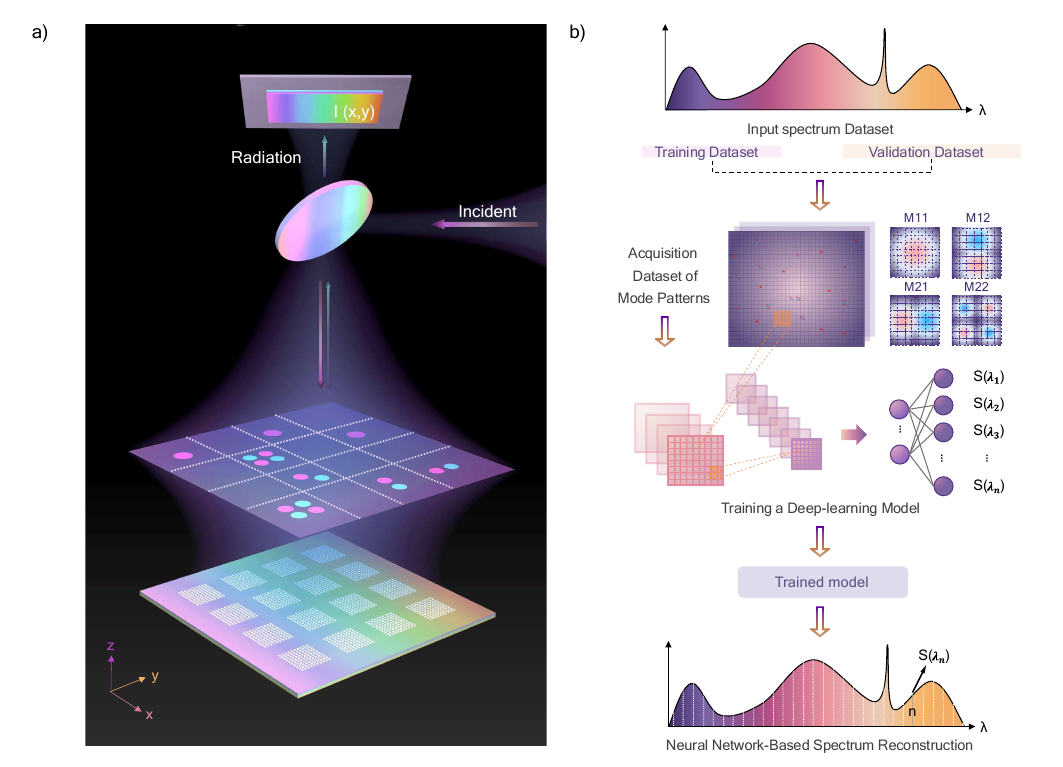}
\caption{\textbf{Schematic and principle of mini-spectrometer.}
(a)
Schematic of mini-spectrometer. The cavities radiate in the vertical direction (blue arrow) under out-of-plane excitation of incident light (purple arrow), each cavity displays a unique radiation pattern when the excitation wavelength matches the cavity resonance. A CMOS camera records the radiation pattern, which contains the spectral characteristics of the incident.
(b)
Principle of end-to-end learning. A large number of spectra with rich features are used as input, and the captured array responses are divided into training and validation datasets. The network establishes an end-to-end connection between the array response and target spectra, mapping the corresponding spectral data to its output.
}
\label{fig1}
\end{figure}

We present the schematic of our mini-spectrometer in Fig. \ref{fig1}(a), in which a large number of radiative high-$Q$ cavities are tightly arranged as a two-dimensional (2D) array. Namely, under out-of-plane excitation of incident light (purple arrows), the cavities radiate toward the vertical direction (blue arrows), and each shows distinct radiative patterns according to the wavelengths and $Q$s of cavity modes. In addition, the radiative patterns are imaged and recorded by the CMOS camera as raw data $I(x,y)$ that contain the spectral characteristics of the incident. This setup is quite simple and compact since neither interconnections between the cavities nor tunable components are needed for scanning.

The raw data is processed using  end-to-end learning techniques (Fig. \ref{fig1}(b)). Specifically, we first feed the spectrometer with a series of pre-defined input spectra with rich spectral features as a training dataset, which could generate distinctive radiative patterns in the vertical emission. By adding noise to the recorded raw images, a deep network with generalization capabilities is trained to map the input spectrum data set $I(x,y)$ to the network output $S(\lambda_n)$, thus establishing an end-to-end connection between the response of the array to the target spectrum without acquiring the detailed attributes of each optical component, which could be quite promising when the scale of the array becomes considerably large. Once the network is trained, the spectrometer can solve arbitrary unknown spectra. Nonvolatile hardware can store the trained network parameters to support portable and in-insu spectral measurement.

\begin{figure}[ht]
\centering\includegraphics[width=18cm]{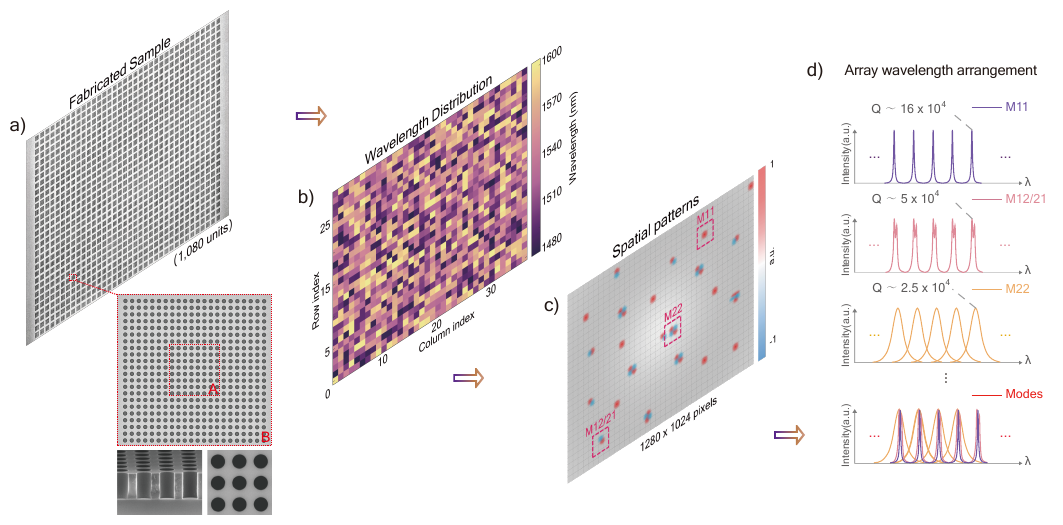}
\caption{\textbf{Fabricated sample and array arrangement strategy.}
(a)
The scanning electron microscope (SEM) images of the fabricated sample show each mini-BIC cavity consisting of region A and region B from top and side views. FIB is used to cleave the PhC for a side view of the etched air holes. 
(b)
Thermogram of wavelength distribution, which contains $30$ rows $\times$ $36$ columns of array cells of $1080$ cavities, aligning with the center wavelength distribution of the M$_{11}$ mode.
(c)
Schematic of spatial patterns when the cavities are excited by the incident.
(d)
The arrangement of the resonance wavelength of cavity modes, in which the high-$Q$ and relatively low-$Q$ modes are intertwined to cover a broad spectrum range.
}
\label{fig2}
\end{figure}

The key to realizing the proposed mini-spectrometer architecture relies on the unique behavior of optical cavities ---  they should possess high-$Q$ factors to ensure high spectral resolution and be capable of out-of-plane excitation to create wavelength-dependent spatial features. The mini-BIC cavity operating within the radiation continuum should be a suitable candidate. As illustrated in Fig. \ref{fig2}, such cavities are fabricated on a silicon-on-insulator (SOI) wafer with a silicon layer thickness of $600$ nm by using electron beam lithography (EBL) and inductively coupled plasma (ICP) dry etching techniques (see Methods for details). The top view shows that the cavity is patterned on a square latticed PhC slab with an air hole radius of $r=175$ nm, which consists of a central region A of size $N_a=8a$ with a periodicity of $a=520$ nm that is surrounded by an outer region B of size $N_b=7b$ with a periodicity of $b=544.5$ nm, resulting in a footprint of $\sim 12\times 12 ~\mu$m$^2$ for a single cavity. The principle of mini-BIC cavities was elaborated in our previous work\cite{2022phcchen}. Briefly, to promote $Q$s, out-of-plane radiation from the cavity is suppressed by a constellation of multiple integer topological charges carried by BICs, while the photonic band gap of the outer region prevents lateral leakage. Owing to topological protection, mini-BIC cavities are quite robust against random fabrication imperfections.  Consequently, the cavities support a set of discrete resonance modes $M_{pq}$ with $Q \sim 10^4-10^5$ that can be readily excited from the vertical direction and exhibit distinctive radiative patterns from far-field observation to distinguish the wavelengths. (See Suppl. Section 1 for details). 

Further, we arrange a large number of mini-BIC cavities to form a 2D array with a margin space of $\sim 5~\mu$m. As shown in Fig. \ref{fig2}(a), the cavity array consists of 30 rows and 36 columns, namely $30 \times 36 =1080$ cavities, which gives a total footprint of $505~\times 606~\mu$m$^2$.  By varying the periodicity $a$ from $520$ to $547$ nm, and accordingly adjusting $b$ from $544.5$ nm to $582.3$ nm, the resonance wavelength of cavity modes shifts, making the array cover a remarkably broad spectral range from $1525-1605$ nm. The thermogram of the simulation wavelength distribution across the array is presented in Fig. \ref{fig2}(b), in which each cavity radiates in featured patterns when the incident wavelength coincides with the cavity modes (Fig. \ref{fig2}(c)). To enrich the features under different wavelength excitations, several strategies are applied during the arrangement of cavities. First, the spacing between individual cavities is set to a sufficient value of $\sim 5~\mu$m to migrate the intercoupling between them, which could avoid the uncertainty caused by fabrication errors. Second, both the fundamental mode $M_{11}$ with the highest $Q$s and higher-order modes such as $M_{12/21}$ and $M_{22}$ with relatively lower $Q$s are adopted to span a wide detection range Fig. \ref{fig2}(d); thus the array can capture the fine and coarse spectral features simultaneously. Numerical simulations (COMSOL Multiphysics) suggest the $Q$s of $M_{11}\sim 16\times10^{4}$, $M_{12/21}\sim 5\times10^{4}$, and $M_{22}\sim 2.5\times10^{4}$ modes, respectively, thus they are expected to distinguish single spectral peak with a minimal bandwidth of 0.01 nm to 0.06 nm. Third, the resonance wavelengths are spatially intertwined across the array, so that the adjacent and nearby cavities can have notable differences in resonance wavelengths to avoid blurring the radiative patterns when the incident spectra contain multiple close sharp peaks. More discussions about the array arrangement are presented in Suppl. Section 2. 

A significant challenge of mini-spectrometers lies in balancing detection resolution and range. Although high-$Q$ cavities can improve spectral resolution, the need for a larger number of cavities to cover a wide spectral range complicates calibration and control. The end-to-end deep learning approach alleviates this problem by employing a cavity array with sufficiently rich and diverse features to differentiate the spectrum, making the specific characteristics of each cavity less critical. Consequently, the end-to-end strategy remains effective despite fabrication imperfections that cause deviations from ideal cavity designs, providing significant tolerance for our proposal.

\section*{Setup and algorithms}

\begin{figure}[ht]
\centering\includegraphics[width=18cm]{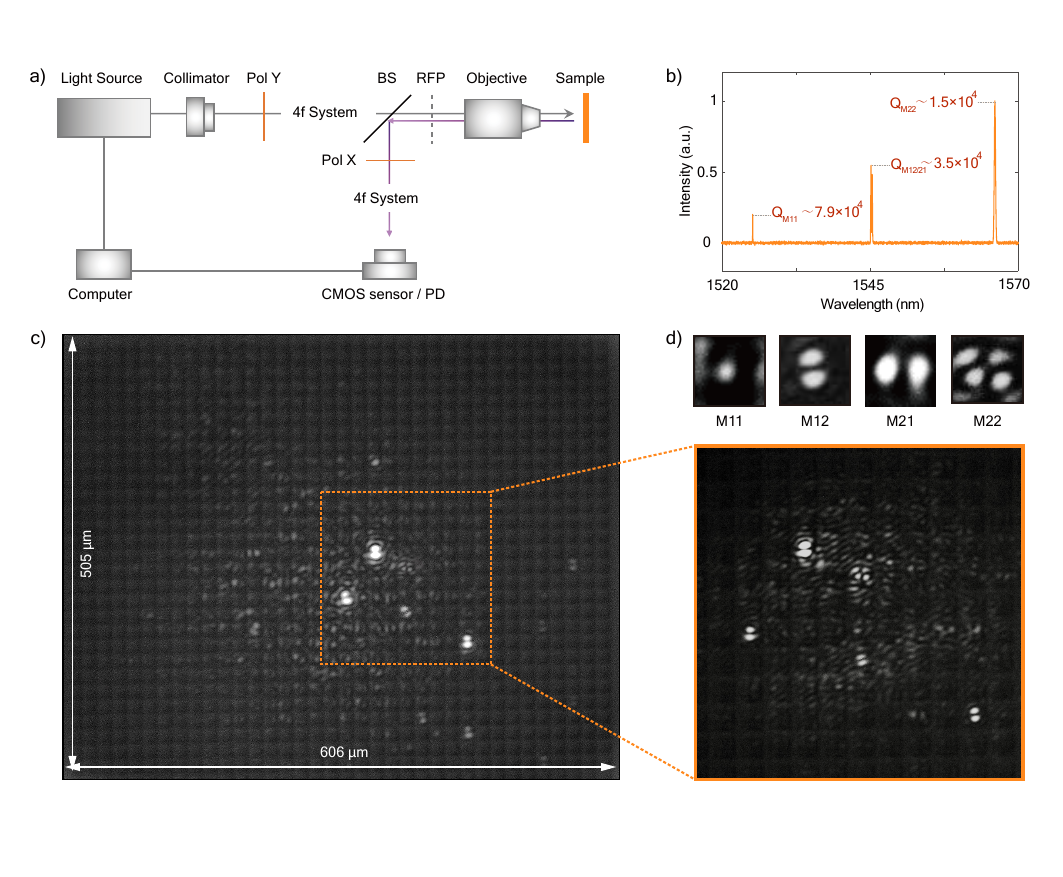}
\caption{\textbf{Experimental setup and results.}
(a) 
Schematic of the measurement setup.  Light Source: an amplified spontaneous emission (ASE) or a narrow line-width laser, BS: beam splitter, RFP: rear focal plane, PD: photodetector.
(b)
Measured reflection spectrum by scanning the wavelength from $1525$ to $1605$ nm using the narrow line-width laser, showing the mini-BIC cavity supports 4 modes in such a range.
(c)
A typical CMOS image illustrates the spatial features of the cavity array when the input wavelength coincides with the resonance.
(d)
A zoom-in view of the CMOS image displays the distinctive pattern of modes $M_{11}$, $M_{12/21}$, and $M_{22}$.
}
\label{fig3}
\end{figure}

Our measurement setup is presented in Fig. \ref{fig3}(a), which is used for different tasks including characterizing the performance of a single cavity, training the end-to-end neural network, and solving unknown spectra. Specifically, incident light from a given light source is collimated and split by a beam splitter (BS) before being focused on the sample with an objective lens (Mitutoyo, NA = 0.26). The reflected light is collected by the same lens, transmitted through a 4$f$ system with a magnification of $0.5$, and detected by either a photodetector (PD, Thorlabs, PDA10PT) or a CMOS camera (ASI990MM-Pro). In this setup, we use cross-filtering techniques\cite{galli2009light} to enhance the signal-to-noise ratios (SNR), by aligning polarizers parallel (Pol Y) and orthogonal (Pol X) to the incident's polarization, which is similar to our previous works \cite{2022phcchen,jin2019topologically,yin2020observation}. The spectra or patterns of reflected light are recorded by the computer for processing and self-adaptive control.

To characterize the cavity's response, a tunable laser light source (Santec TSL-550) is used to sweep with high precision at a single wavelength in the range of $1525$ to $1605$ nm, with a fixed power of 12 dBm. The measured reflectivity and $Q$s of a single cavity are presented in the upper panel of Fig. \ref{fig3}(b). We identified four modes: M$_{11}$, M$_{12/21}$, and M$_{22}$, demonstrating the quantization of cavity modes due to the lateral reflective boundary. The $Q$s of these modes, extracted from Lorentzian-fitted spectra, show that the M$_{11}$ reaches $Q\sim7.9\times 10^{4}$. On the same sample, the $Q$s of M$_{12}$ and M$_{22}$ are obtained as $3.5\times10^{4}$ and $1.5\times10^{4}$, respectively. It is worth noting that, since the underlying SiO$_2$ layer is not removed for mechanical stability, the $Q$s of the cavities are slightly lower than previously reported results, as the up-down mirror symmetry is broken. When the incident wavelength aligns with the cavity modes, distinct radiative patterns emerge, which can serve as spatial features to differentiate the spectra.

To measure the response of the cavity array, we use the same setup with a different light source and detector. We employ an amplified spontaneous emission (ASE, Beogold, BG-ASE-SS-CL) light source operating in the C+L-band to generate a broad spectrum, which is further shaped by a programmable waveform shaper (Finisar, WaveShaper 4000S) to create arbitrary spectra, and the spatial patterns of cavity array are recorded by the CMOS camera in a resolution of $1280 \times 1024$ pixels. The waveform shaper and the CMOS camera are controlled by a computer that synchronizes the spectrum shaping and data acquisition. This setup can automatically generate and capture the dataset for both neural network training and validation. For better repeatability, the setup is placed in a room environment with a controlled ambient temperature of 24$^\circ C$. A typical CMOS image of the entire cavity array and a zoomed-in section are shown in Fig. \ref{fig3}(c) and (d), clearly illustrating the spatial features of the radiative pattern under the excitation of a specific spectrum --- a few cavities become bright when the spectrum overlaps with the resonant wavelength, appearing as a combination of individual cavity mode patterns.

\begin{figure}[ht]
\centering\includegraphics[width=18cm]{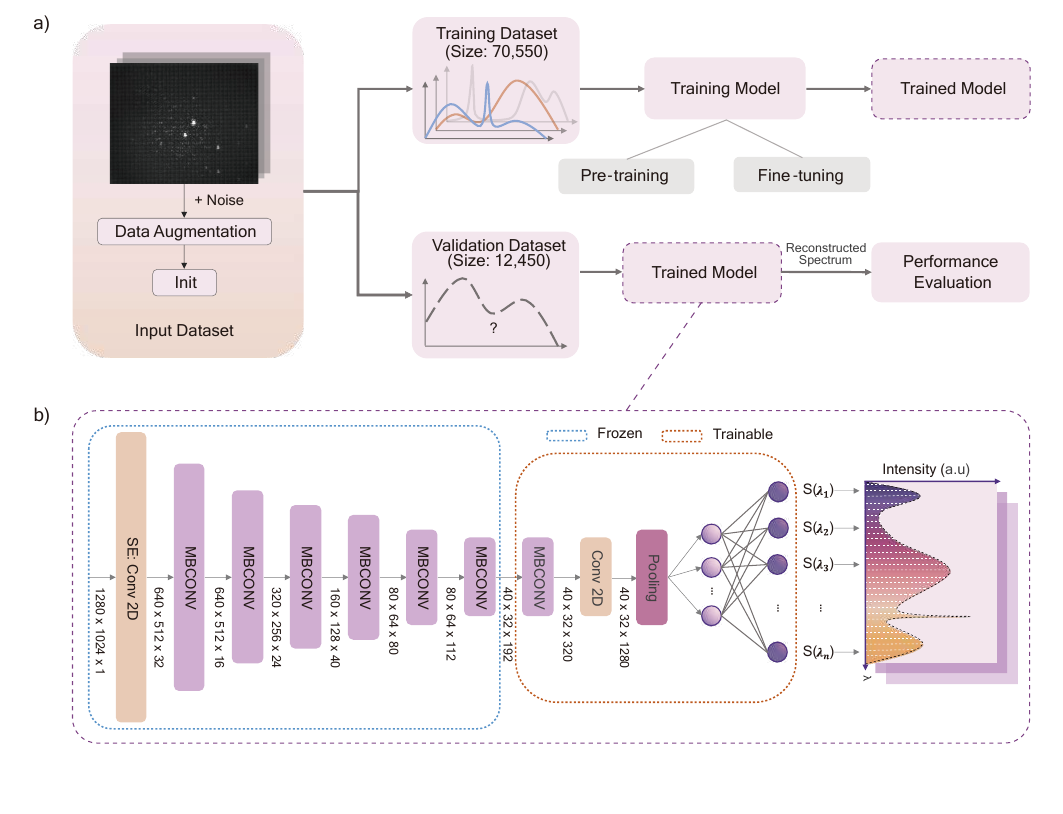}
\caption{\textbf{End-to-end training process network.}
(a)
The generalization ability of the deep-learning model is enhanced by introducing trace noise to 40\% of the dataset (83,000 images) before feeding it into the network. The dataset is split into training and validation sets. The model is trained through pre-training and fine-tuning on the training set, and the reconstructed spectrum is evaluated using the validation set, which was not involved in the training process.
(b)
Detailed EfficientNet model with the network output corresponding to the spectra. The yellow dashed box represents the trainable layers involved in the full training process, while the blue dashed box indicates the frozen layers, which remain unchanged during pre-training.
}
\label{fig4}
\end{figure}

The captured CMOS image is utilized to train and validate the end-to-end neural network for spectroscopy. The detailed processing framework is illustrated in Fig. \ref{fig4}. Initially, the captured images, with a resolution of $1280 \times 1024$ pixels, are divided into a $36 \times 30$ grids to correspond with the cavity array scale, ensuring that each grid cell contains a single cavity. Each cell is automatically segmented into $34 \times 34$ pixels to accurately recognize the spatial patterns. To enhance the neural network’s generalization ability, 2\% Gaussian noise is added to the raw data as a common technique in deep learning. These processed images serve as input to the neural network for further steps.

Next, we generate $83,000$ random spectra using the waveform shaper as the input dataset, with $70,550$ (85\%) used for training and $12,450$ (15\%) for validation. A two-stage transfer learning strategy was applied during model training: pre-training and fine-tuning. At the first stage, 65\% of the training dataset was used for pre-training, where certain network layers were frozen (indicated by the blue dashed box in Fig. \ref{fig4}(b)), while only the trainable layers (orange dashed box) were updated. After pre-training, all layers were unfrozen and the remaining 35\% of the training data were used to further optimize the entire network, referred to as the fine-tuning stage. This approach reduces the computational cost since only a subset of layers are involved during pre-training thus effectively speeding up the process.

Our end-to-end network is built on the EfficientNet architecture, which is commonly used for deep learning tasks due to its remarkable ability to reduce computational costs while ensuring accuracy in presenting the targeted features\cite{koonce2021efficientnet}.  EfficientNet applies optimized convolutional neural networks (CNNs) with a compound scaling strategy. The architecture combines several key components, including a standard convolutional layer, a mobile inverted bottleneck convolution (MBConv) layer, a squeeze-and-excitation (SE) module, and a fully connected layer, to efficiently manage complex spectral reconstruction tasks. In the output layer, we employ $8,000$ nodes to present the light intensities ranging from $1525$ to $1605$ nm with a spectral resolution of $0.01$ nm, which fits well with the distinguishing capability of high-$Q$ mini-BIC cavities for spectroscopy. The network can capture both local and global features in the spectral data since the MBConv layer particularly addresses feature extraction through depth-wise separable convolution and expansion layers. Therefore, our deep learning architecture can reduce the number of required parameters, thus providing significant computational efficiency for processing spectral data.

Besides, we employ migration learning techniques to fine-tune EfficientNet, which further promotes the accuracy and stability of spectral reconstruction by leveraging generic features learned from pre-trained models on large-scale datasets. This approach is particularly suitable for the size of our dataset ($8,000$ data output points), effectively reducing the risk of overfitting that can occur with small data volumes and speeding up model convergence. The model is trained using a composite loss function that combines mean square error (MSE) and cosine similarity to ensure precise reconstruction of the spectral signal features. To evaluate the quality of the reconstruction, we also computed the fidelity between the reconstructed and reference spectra, ensuring a high level of agreement between the output and the actual spectrum.
More details about the deep learning network architecture and algorithms are presented in Suppl. Section 3.

\section*{Experimental results}
To validate the proposed mini-spectrometer, we conduct a series of experiments demonstrating its ability to reconstruct arbitrary spectra within the detection range of $1525–1605$ nm, namely within a broad bandwidth of $80$ nm. For comparison, the same spectra were also measured by using a commercial spectrometer (YOKOGAWA AQ6374) with a spectral resolution of $0.02$ nm to serve as reference values. The outputs of our mini-spectrometer and commercial spectrometer are directly compared for blind testing. 

We first characterized several typical spectrum detection scenarios, including slow-varying features, sharp peaks, and their combinations. As shown in Fig. \ref{fig5}(a), the reference data reveals a series of spectra with slow-varying envelopes across the detection range, featuring gentle peaks at either the lower or higher wavelength ends. The outputs from our mini-spectrometer match well with the reference values, confirming its capability for broadband detection. Additionally, we generated a set of spectra containing multiple narrow-band peaks with varying spectral spacings and intensities (Fig. \ref{fig5}(b)). Our mini-spectrometer accurately identified these fine features, correctly pinpointing peak positions and showing similar intensity levels when compared to the reference data. Moreover, the mini-spectrometer was able to solve spectra combining both coarse and fine features (Fig. \ref{fig5}(c)), further validating its effectiveness in detecting a variety of spectral features across a broadband range. 

We further evaluate the spectral resolution of our mini-spectrometer through additional tests. First, we attempted to detect a single sharp peak with the full width at half maximum (FWHM) of $0.04$ nm. As shown in Fig. \ref{fig5}(d), our mini-spectrometer successfully captured this ultra-fine spectral feature, measuring the FWHM of $0.048$ nm, which closely aligns with the reference value. Due to the accuracy limit of the trained model, some small ripples with normalized intensities of $0.03$ appear on the output of our mini-spectrometer.  Additionally, we tested the spectrometer's ability to distinguish two adjacent narrow peaks, each with the FWHM of $0.04$ nm and spaced $0.08$ nm apart. The measured result in Fig. \ref{fig5}(e) demonstrates that our mini-spectrometer correctly reconstructs the two peaks. Based on these results, we conclude that our miniature spectrometer achieves a resolution of $0.048$ nm in terms of distinguishing minimal spectral peaks.


To confirm that the spectral resolution remains consistent across the entire detection range, we further carry out a test of detecting a single ultra-narrow peak appearing at any position between $1525$ and $1605$ nm. We generated a set of spectra, each containing a single peak with the FWHM of $0.04$ nm, and then shifted their peak positions in $0.1$ nm increments to cover the full detection range. Using our mini-spectrometer, we successfully solved all these spectra, showing excellent agreement between the measured and reference values. The peak deviation had a mean value of $0.008$ nm across the detection range (Fig. \ref{fig5}(f)), with calculated MSE error below 0.006 parts per thousand for linear fitting the wavelength sweeping. Additionally, we introduced the term ``FWHM error" to represent the difference between the measured and nominal FWHM values, serving as an indicator of detection accuracy. The standard deviation of the FWHM error for our mini-spectrometer was calculated to be $0.027$ nm, which is larger than that of a commercial spectrometer with an FWHM error deviation of $0.005$ nm, but still a fairly good value for on-chip spectroscopy. 

To statistically show the effectiveness of our mini-spectrometer, we analyze the fidelity of our mini-spectrometer in detecting 12,450 blind spectra by using a Bootstrap method\cite{davison1997bootstrap}. The Bootstrap method conducts extensive random sampling of the fidelity of blindly measured spectra, summarizing the results of all samples. It calculates the mean values, standard deviations, and confidence levels to evaluate the fidelity distribution of the reconstructed spectra without making prior assumptions about the data distribution, thus enabling an effective assessment of the mini-spectrometer's performance of multi-featured spectra. The resulting histogram is presented in Fig. \ref{fig5}(g), showing that the mean fidelity value exceeds $95\%$ with a standard deviation of $0.27\%$, which are applicable values for portable diagnosis.

At last, we evaluate the sensitivity of our mini-spectrometer, namely the minimum power required by the spectrometer to correctly solve the spectra. The sensitivity of the spectrometer depends on the sensitivity of the detector and the insertion loss of optical components. We assume the detector operates in the linear region that pins a minimum detectable power. The insertion loss of the entire optical setup is measured as $\sim 14$ dB, which includes the losses of the planar and objective lenses, polarizers, beam splitter, and optical cavities. Unlike the micro-ring-based mini-spectrometer architecture,  our design is excited from the out-of-plane direction and thus doesn't require the coupling, distributing, and routing of light in the transverse direction, which can significantly reduce the insertion loss to promote sensitivity that avoids the usage of the extra optical amplifier. We also take a single narrow peak with FWHM of $0.048$ nm to characterize the detection sensitivity of our mini-spectrometer. The narrow-peak feature was accurately solved at a low incident power of $0$ dBm, corresponding to a detection sensitivity of  $12.5~\mu$W/nm. More data about the experiment results are presented in Suppl. Section. 4.

\section*{Discussion}
Our implementation of the mini-spectrometer upon a large-scale radiative cavity array demonstrates that parallel detection, combined with high-$Q$ optical cavities in the continuum, effectively addresses the trade-off between detection range and resolution for on-chip spectroscopy. Using the array architecture, the detection range can be easily extended to cover a wide spectrum, avoiding the need to control or scan a single dispersive component, which remarkably simplifies the detection process. Furthermore, the high-$Q$ optical cavities enable the differentiation of minor spectral differences, thereby providing a high spectral resolution. The mini-BIC cavity can be excited from the out-of-plane direction, which reduces the insertion loss for on-chip routing and distributing the light, offering lower system complexity and better detection sensitivity. 

The challenge of parallel detection using the array architecture lies in the participation of a large number of dispersive components, making it nearly impossible to accurately fabricate and calibrate each component. The end-to-end deep learning strategy offers a viable solution to this issue\cite{peng2022end,gupta2024senglean}. When optical cavities exhibit varying behaviors under the excitation of different wavelengths, their ability to fully cover all possible spectral features becomes more important than precisely determining individual responses. Consequently, the deep learning network can capture these features without needing predefined knowledge about each cavity by applying the end-to-end training strategy\cite{signoroni2019deep,gao2022computational,zhang2020solver}.

Our mini-spectrometer achieves a detection range of $80$ nm and a resolution of $0.048$ nm. In our experiments, the detection range is limited by the spectral range available for generating arbitrary waveforms for training, namely the limited bandwidth of the ASE light source and waveshaper; however, it can be easily extended with additional broadband training data. Furthermore, our cavity operates at $Q\sim 6.9 \times 10^4$, which determines the minimal distinguishable feature in the spectrum. Increasing $Q$s can effectively promote spectral resolution but can reduce detection sensitivity as the excitation of cavities becomes less effective. On the other hand, the high-$Q$ of characteristics of microcavities can notably enhance the strength of light-matter interaction, paving the way for combining wavelength-selective materials such as black phosphorus or quantum dots for more sensitive spectral detection. We argue that our mini-BIC array architecture is scalable and can be flexibly tailored for different application scenarios. More discussion about promoting the performance and comparison of reported mini-spectrometers is presented in  Suppl. Section. 5.

\section*{Conclusion}
To sum up, we propose and demonstrate a compact, high-resolution, broadband miniaturized spectrometer by combining a large-scale array of radiative mini-BIC cavities with an end-to-end deep learning network. Our method avoids the usage of scanning components which are usually complex and require long-time acquisition and precise alignment, thus solving the dilemma of calibrating a large number of dispersive components in conventional array-based mini-spectrometers.  
Our mini-spectrometer achieves a detection range of $80$ nm, resolution of $0.048$ nm, and sensitivity of $12.5~\mu$W/nm on a footprint of $505~\mu$m $\times$ $606~\mu$m. The proposed architecture is scalable to fulfill the requirement of a broader detection range and higher resolution, thus paving a promising way for high-performance on-chip spectroscopy for spectral diagnostics for many chemical, biological, and medical applications.

\section*{Methods}

\section*{\underline{Numerical simulations}}
All simulations are performed using the COMSOL Multiphysics in the frequency domain. Three-dimensional models are created with photonic crystal slabs between two perfect matching layers (PML) in the $z$ direction. For unit cell simulations, periodic boundary conditions are adopted in the $x$ and $y$ directions. For cavity simulations, PMLs are applied in the transverse directions. The spatial meshing resolution is adjusted to adequately capture resonances with $Q$s of up to $10^{5}$. The eigenvalue solver is used to compute the frequencies and the quality factors of the resonances, as shown in Fig. \ref{fig2}d. The mode radiation patterns are calculated by  retrieving the complex electric fields $E_{0,j}(j=x, y)$ right above the PhC surface and then followed by the diffraction integral as\cite{liang2012three}:

\begin{align}
F_{j}\left(\theta ,\phi \right) \propto \left(cos\theta  + cos\phi  - 1\right)\iint _{x,y}^{}E_{0,j}\left(x,y\right)e^{ - ik_{0}\left(tan\theta _{x} + tan\phi _{y}\right)}dxdy.
\end{align}

\section*{\underline{Sample fabrication}}
We fabricate the sample on a silicon-on-insulator (SOI) wafer with e-beam lithography (EBL) and induced coupled plasma (ICP) etching. For the EBL process, we first spin-coat the cleaved SOI chips with a $420$ nm-thickness of ZEP520A photo-resist and then expose it with EBL (Elionix, ELS-F125) at a beam current of $1$ nA and a field size of $500$ $\mu$m. Then we etch the sample with ICP (Oxford Plasmapro Estrelas 100) using a mixture of SF$_{6}$ and CHF$_{3}$. After etching, we remove the resist with Dimethylacetamide (DMAC).

\section*{\underline{Measurement}}

We use a tunable laser (Santec TSL-550, C + L band) to generate incident light. The light is first sent through a polarizer (Y-Pol) and is focused by a lens (L1) onto the rear focal plane of an objective (Mitutoyo Apo NIR, 50$\times$). The reflected and scattered light is collected by the same objective and a 4f system is used to adjust the magnification ratio to 0.5× to best fit the observation. After passing through an orthogonal polarizer (X-Pol), only the scattered light is collected using a photo-diode (PDA10DT-EC). The resonance peaks are recorded by a high-speed data acquisition card (NI PCIe-6361) connected to the photo-diode during wavelength scanning and then fitted to the Lorentzian function. A flip mirror is used to switch between the InGaAs infrared CMOS camera (ASI990MM-Pro) and photodetector.

We switched the objective lens to 5$\times$ and used an amplified spontaneous emission (ASE, Beogold, BG-ASE-SS-CL) light source operating in the C+L band to produce a broad spectrum, which was further shaped by a programmable waveform shaper (Finisar, WaveShaper 4000S) to generate an arbitrary spectrum. The spatial pattern of the cavity array was recorded by a CMOS camera at a resolution of 1280 × 1024 pixels. The waveshaper and CMOS camera are controlled by a computer, which synchronizes spectrum shaping and data acquisition. The measurement system automatically generates and captures data sets for neural network training and validation.

\section*{\underline{Data processing}}

The captured images were used to train and validate an end-to-end neural network for spectral analysis. After accounting for inherent CMOS noise and environmental background noise, 40\% of the acquired images were augmented with 2\% Gaussian noise. These processed images were used as input to the neural network. The spectral reconstruction network is based on the EfficientNet-B0 architecture, where the input consists of grayscale images with a resolution of $1280 \times 1024 \times 1$. In the stem layer, a $3 \times 3$ convolution with a stride of 2 downsamples the input to $640 \times 512 \times 32$, extracting initial spatial features. The subsequent MBConv layers further optimize the network structure for efficiency. The first stage uses MBConv1, maintaining the number of channels at 32 but reducing the resolution to $640 \times 512 \times 16$ via a $3 \times 3$ convolution. The second stage employs MBConv6 (stride of 2) to produce a feature map of $320 \times 256 \times 24$. The third stage uses a $5 \times 5$ convolution kernel to downsample to $160 \times 128 \times 40$, followed by the fourth stage, which reduces the feature map to $80 \times 64 \times 80$. The fifth and sixth stages both use $5 \times 5$ convolutions, increasing the number of channels to 112 and 192, respectively, while adjusting the spatial resolution. The seventh stage maintains the same resolution while increasing the number of channels to 320. Subsequently, a $1 \times 1$ convolution expands the channel count to 1280, and global average pooling generates a 1280-dimensional vector. Finally, a fully connected layer outputs $8000$ nodes to accommodate the spectral reconstruction task, which spans a wavelength range of $80$ nm with $0.01$ nm accuracy. In addition, a two-stage migration learning strategy was used to enhance model training. Firstly, pre-training was performed using 65\% of the training data, updating only the network parameters of Pooling, Conv2, and the last layer MBConv, while the other network layers were kept frozen. At the end of the pre-training, all layers were unfrozen and the remaining 35\% of the data was used to fine-tune the model.

The network uses the Swish activation function, and batch normalization along with dropout (dropout rate of 30\%) is applied after each convolutional and fully connected layer for regularization to prevent overfitting. During training, the AdamW optimizer is used, with an initial learning rate of $1 \times 10^{-4}$, and learning rate decay is applied when performance on the validation set plateaus. The loss function is a composite of MSE and cosine similarity, evaluated over $8000$ wavelength points between the network output and ground truth spectrum. All image preprocessing, network training and testing were implemented in Python using OpenCV and PyTorch libraries, with training and inference performed on a cloud server equipped with NVIDIA L20 GPUs.

\newpage
\section*{References}
\bibliography{mainbib.bib}
\bibliographystyle{naturemag}

\clearpage

\clearpage
\begin{figure}[htbp] 
 \centering 
 \includegraphics[width=18cm]{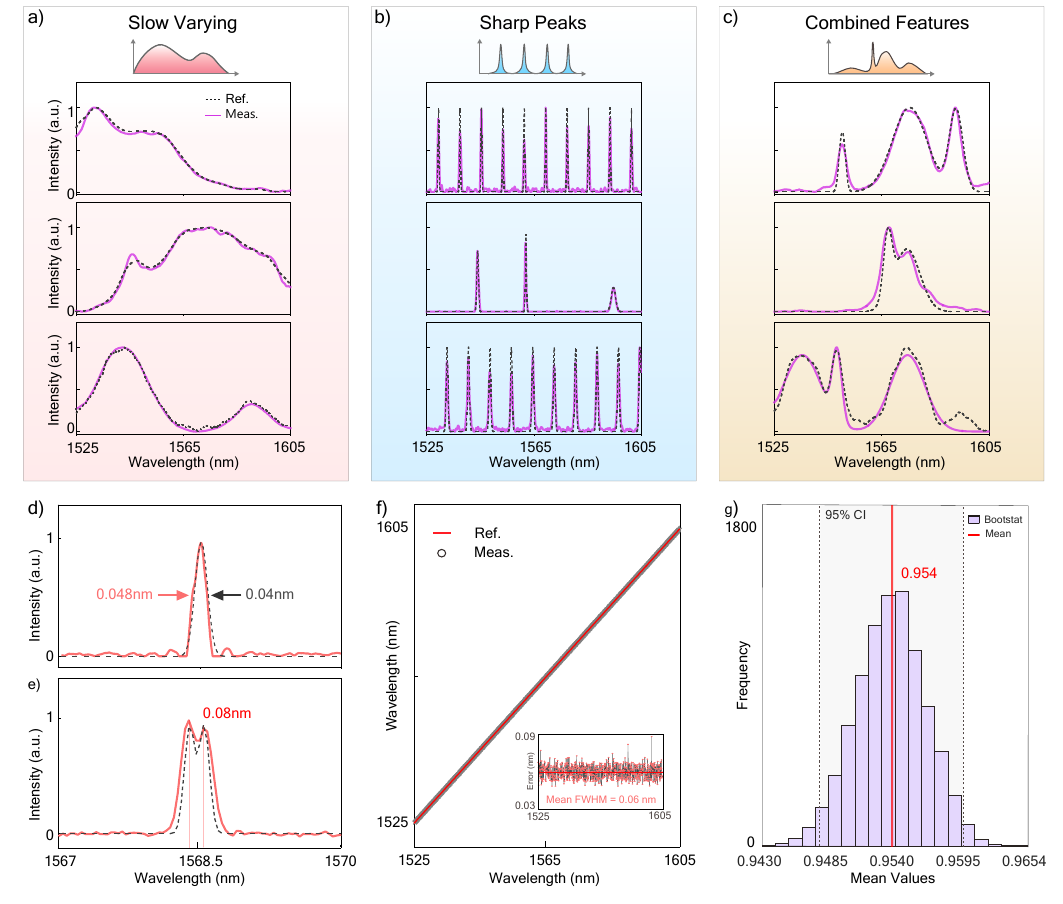}
\end{figure}

\begin{figure}[htbp] 
\caption{\textbf{Blind tests of spectral reconstruction.} The outputs of our mini-spectrometer are compared with a commercial spectrometer as reference values for a series of representative spectral features. 
The reconstructed spectra (purple solid line) agree well with the reference values (black dashed line), for
(a)
Slow varying features
(b)
Sharp peaks features.
(c)
Combined features.
(d) 
Reconstruction of a single sharp peak with an FWHM of $0.048$ nm.
(e) 
Reconstruction of dual peaks with FWHMs of $0.048$ nm separating in $0.08$ nm.
(f)
Evaluation of reconstructing 801 sharp peaks from $1525$ to $1605$ nm with a step increment of $0.1$ nm. The inset shows the mean FWHM error compared with the reference values as ground truth.
(g)
The histogram of reconstruction fidelity for $12,450$ blind spectra using the Bootstrap method shows that the mean fidelity of our mini-spectrometer exceeds $95\%$.
}
\label{fig5}
\end{figure}

\end{document}